\documentstyle[aps,multicol,epsf]{revtex}
\tightenlines

\include{psfig}

\begin{document}
\draft 

\title{Applications of the Stell-Hemmer Potential to Understanding
Second Critical Points in Real Systems}

\author{A. Scala, M. R. Sadr-Lahijany, N. Giovambattista,
S. V. Buldyrev, H. E. Stanley}

\address{Center for Polymer Studies and Department of Physics\\
        Boston University, Boston, Massachusetts 02215.}

\date{\today}

\maketitle


\begin{abstract}

We consider the novel properties of the Stell-Hemmer core-softened
potentials.  First we explore how the theoretically predicted second
critical point for these potentials is related to the occurrence of
the experimentally observed solid-solid isostructural critical
point. We then discuss how this class of potentials can generate
anomalies analogous to those found experimentally in liquid water.

\end{abstract}

\pacs{PACS numbers: 61.20.Gy, 61.25.Em, 65.70.+y, 64.70.Ja}

\begin{multicols}{2}

\begin{figure}[htbp]
\begin{center}
\mbox{\psfig{figure=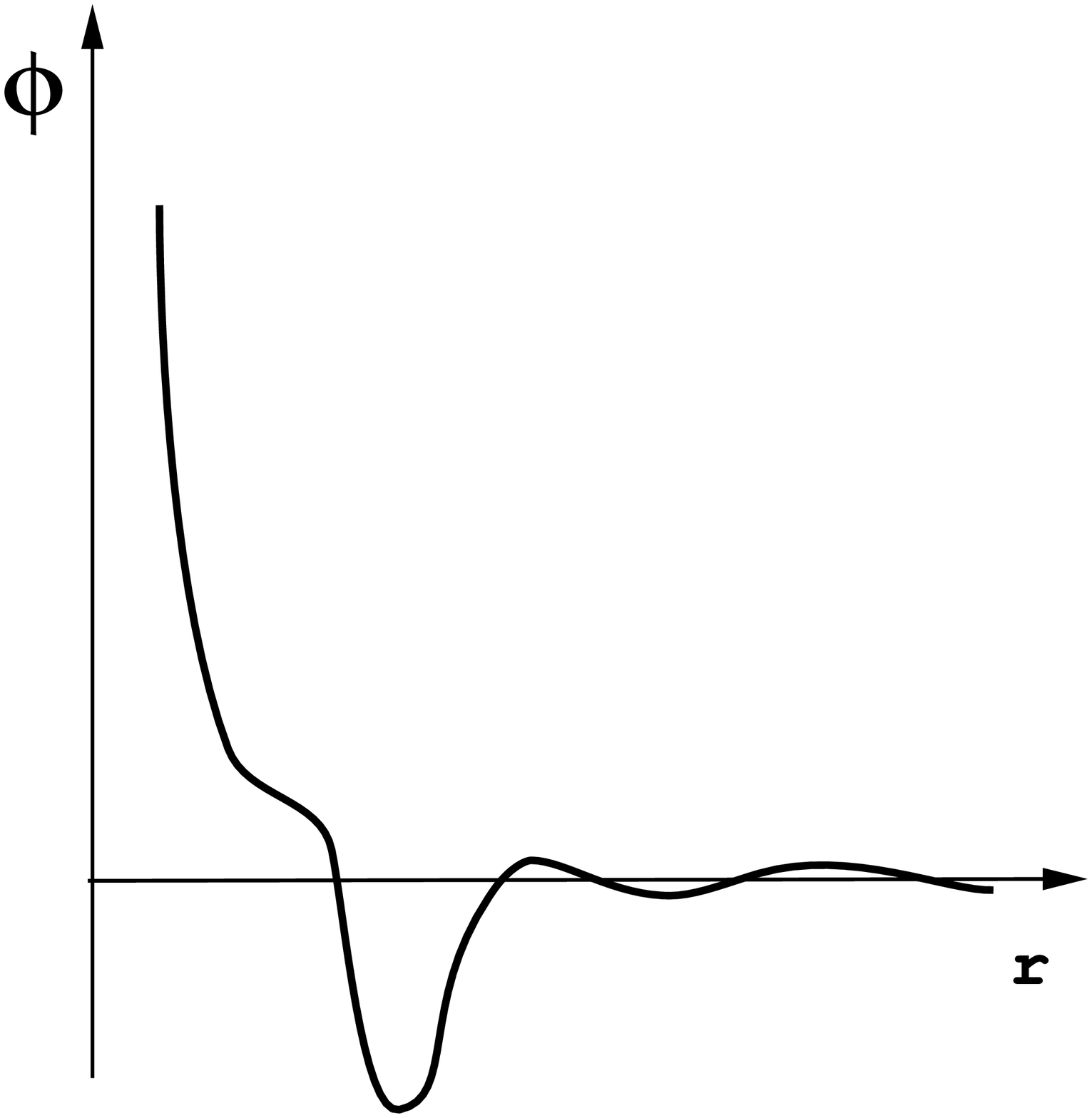,width=15cm,angle=0}}
\end{center}
\caption{Example of core-softened potential (the soft core is a region
where the potential is concave).}
\label{fig:ShoulderPotential}
\end{figure}

Simple liquids are often modeled by pairwise potentials possessing a
repulsive core, mimicking the impenetrability of atoms or molecules,
and an attractive part, responsible for the gas-liquid transition.
For a fluid in which the interaction potential $\phi (r)$ has a hard
core plus an attractive part, a ``softened'' hard
core~(Fig.~\ref{fig:ShoulderPotential}) can produce an additional line
of phase transition~\cite{Stell1}. A general argument by Stell and
Hemmer based on the symmetry between occupied and unoccupied cells in
lattice gas models of fluids predicts that the the additional line of
phase transition can end at a novel critical
point~\cite{Stell1,Rowlinson}. An explicit example of the occurrence
of a second line of first order transition is given in~\cite{Stell1}
for a one-dimensional continuum model of a fluid with long-range
attraction~\cite{Kac}; this first order line can end in a critical
point depending on the details of the core-softened potential.  The
general result is that if the repulsive part (core) of the interaction
potential has a concave part (which makes it ``core-softened''), then
it is likely that such a novel transition occurs~\cite{Stell3}.  Stell
and Hemmer~\cite{Stell1} relate the occurrence of a high-density,
low-temperature critical point to the known isostructural solid-solid
critical point observed in some experiments~\cite{Jayaramen}.

\begin{figure}[htbp]
\begin{center}
\mbox{\psfig{figure=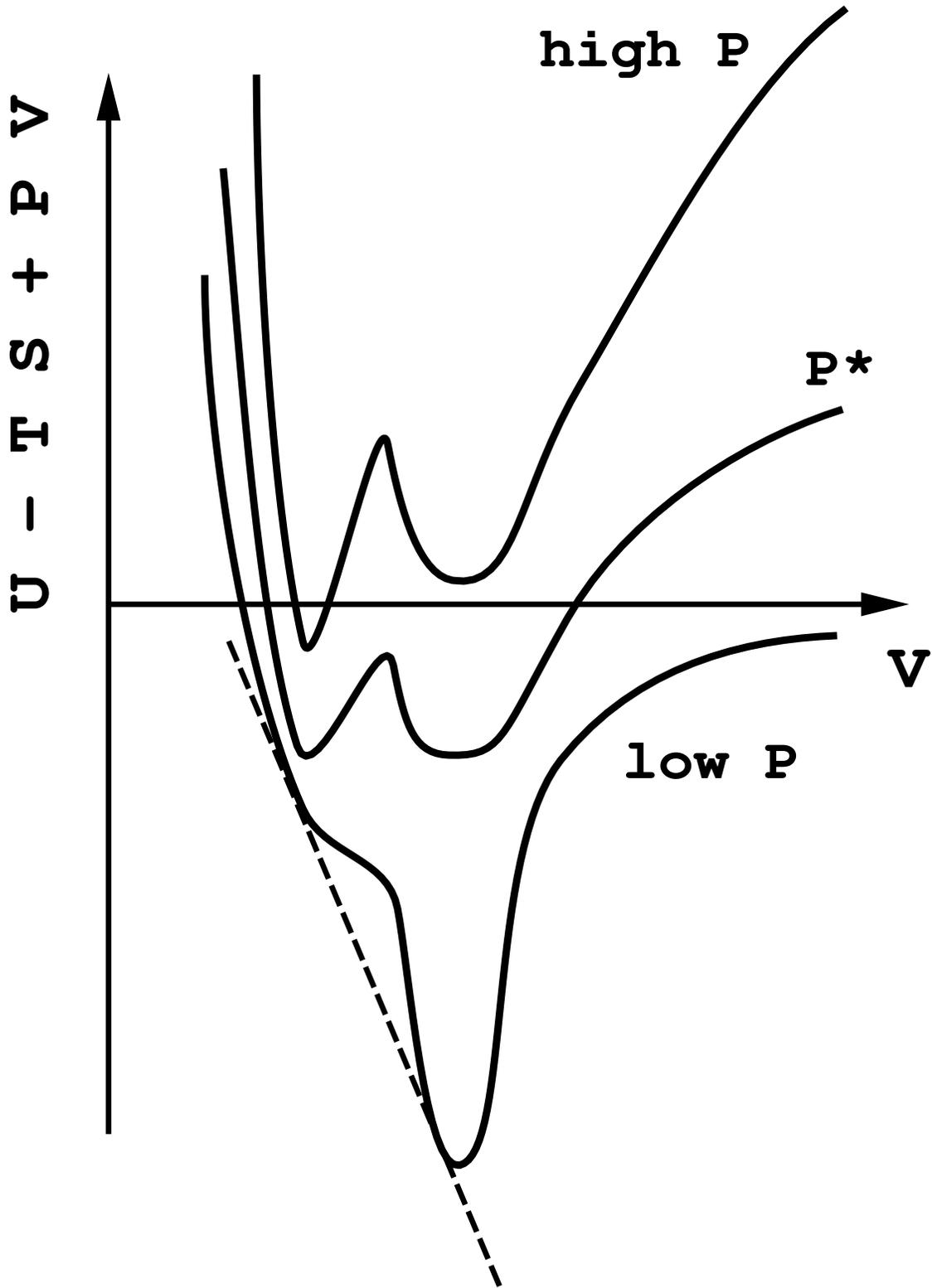,width=15cm,angle=0}}
\end{center}
\caption{The shape of $U+PV-TS$ at $T=0$ for various $P$. Note the
evolution of the minima with pressures: at high enough pressure the
absolute minimum is at higher densities than the absolute minimum at
low pressures, signaling a first order transition. The transition
occurs at the pressure $P^*$ equal to the absolute value of the slope
of the long-dashed line.}
\label{fig:FreeEnergy}
\end{figure}

To understand the occurrence of a second transition, consider the
Gibbs potential at zero temperature.  The shape of the energy $U$ as a
function of the volume $V$ should have the same ``core-softened''
shape (i.e. possesses a region where it is concave) as the
inter particle potential.  The stable phase is then determined by the
Gibbs potential
\begin{equation}
\label{e1x}
G(P,T)\equiv\mathop{\mbox{min}}_V\{U+PV-TS\},
\end{equation}
where $S$ is the entropy. The right hand side of~(\ref{e1x}) is shown
in~Fig.~\ref{fig:FreeEnergy} as a function of $V$, for $T=0$ and for
different values of pressure $P$. At low $P$,
the stable phase of the system has a specific volume at which the
average inter-particle distance is near to the minimum of the
inter-particle potential.  The concavity in $U$ assures that, on
increasing $P$, an additional minimum appears in $U+PV-TS$.  For high
enough $P$, this minimum will become the lowest one and the stable
phase of the system will be the one for which the mean inter-particle
distance is ``inside'' the softened part of the core.

For one-dimensional models, a first-order transition at $T=0$ between
a dense and an open phase occurs at a pressure that can be determined
exactly with the graphical construction
of~Fig.~\ref{fig:FreeEnergy}~\cite{Yoshimura,Sadr}.

For short-range interactions, the entropy gives a huge contribution in
one dimension, making any phase transition disappear for $T>0$. Hence,
the contribution of the entropic term $TS$ makes the double well
structure of~Fig.~\ref{fig:FreeEnergy} disappear when $T>0$. This may
not be true in higher dimensions, and so a line of first order
transitions (eventually ending in a critical point) could be present
for $T>0$.

Stronger evidence for the occurrence of a high-density,
low-temperature critical point is given in~\cite{Stell4} upon
developing and refining analytic methods to investigate the
high-density region of the phase diagram of a fluid. The methods used
are in principle for a dense fluid, and hence would predict a
liquid-liquid transition. However, the second critical point can be
related to the isostructural critical point occurring in the solid
phase of materials such as Cs,Ce and of mixtures such as Sm-S and
Ce-Th~\cite{ExperIsostrCritPoint}. For all these materials the shape
of the effective pair potential is
``core-softened''~\cite{ExperCoreSoft}.

\begin{figure}[htbp]
\begin{center}
\mbox{\psfig{figure=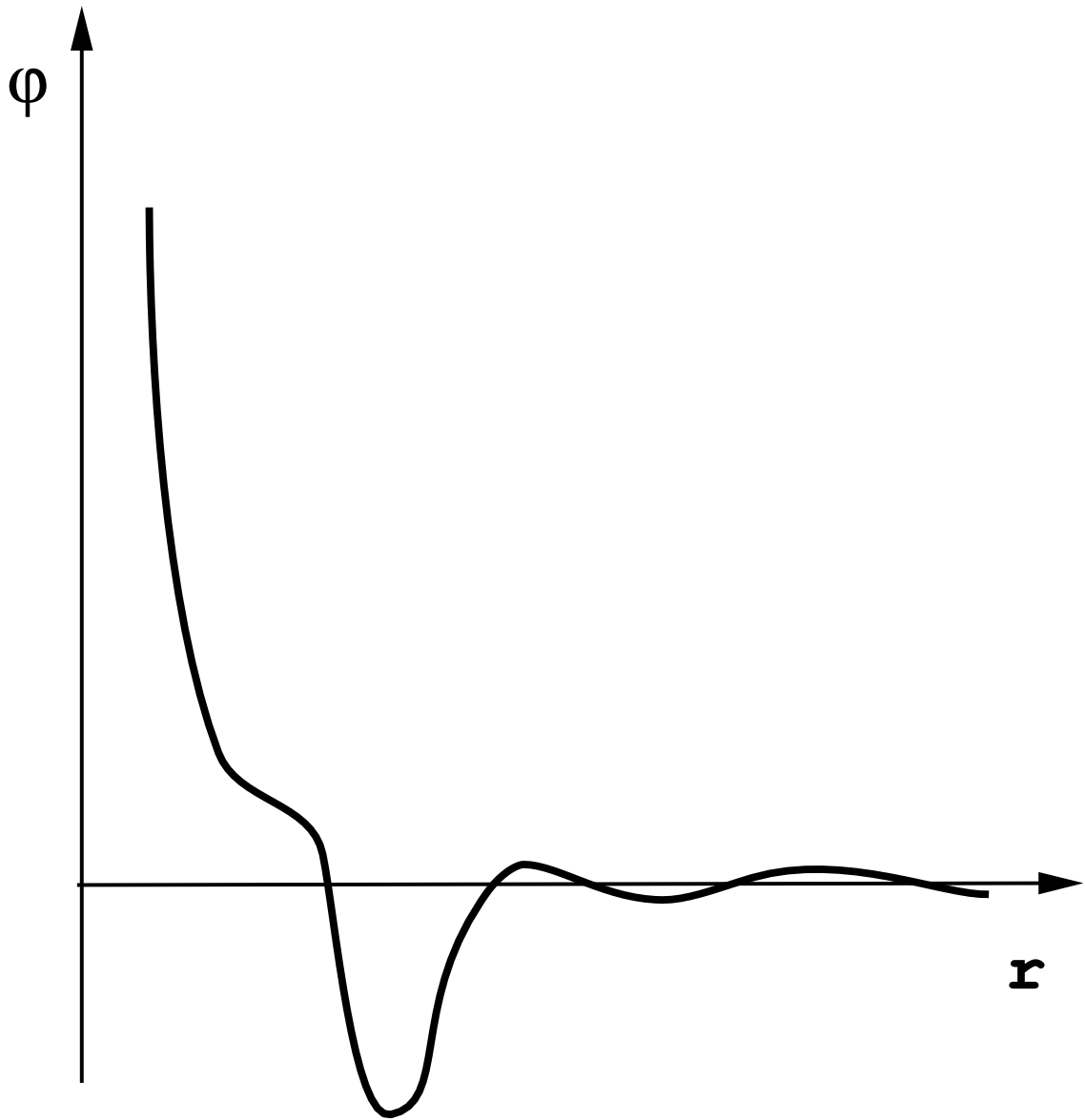,width=15cm,angle=0}}
\end{center}
\caption{Schematic form of the potential as obtained from the
inversion of scattering data or from first principle calculations. See,
e.g.,~\protect\cite{InvertMetal,InvertWater}.}
\label{fig:InvertPotential}
\end{figure}

Many liquid metals (Ga and Sn are prominent examples) have static
structure factors ${\cal S}(k)$ that show weak subsidiary maxima, or
asymmetries, in the main peak of ${\cal S}(k)$ that suggest the
presence of a ``structured'' core that is not infinitely
steep. First-principle calculation of the effective ion-ion potential
for Ga leads to a core-softened potential~\cite{CalculateMetal}. Monte
Carlo simulations with this potential reproduce for Ga the observed
anomalies in ${\cal S}(k)$. Inversion of the experimental structure factors
for In, Zn, Al, Ge, Sn, Cs, Rb, Tl, and Pb, using random phase
approximation or the Ornstein-Zernike equation with a closure, also
results in effective core-softened
potentials~(Fig.~\ref{fig:InvertPotential})~\cite{InvertMetal}.

In addition to the second critical point, core-softened potentials can
produce a density anomaly, i.e.  the material can expand upon
cooling. The occurrence of crossing isotherms was observed in the
model~\cite{Stell2}. It was noted that, although isotherms crossing is
not a common feature in fluids, it is to be found whenever $(\partial
V/\partial T)_P$ changes sign, as it does in water at approximately
4$^\circ$C and atmospheric pressure. Using thermodynamic arguments,
Debenedetti et al. also noted that a ``softened core'' can cause the
thermal expansion coefficient $\alpha_P \equiv (1/V) (\partial V /
\partial T)_P $ to be negative~\cite{Debenedetti}.
 
One-dimensional fluids~\cite{Takahashi} with core-softened potentials
have been studied in relation to the molecular origin of the negative
thermal expansion in two fluids, water and tellurium~\cite{Yoshimura},
which have an effective core-softened
potential~\cite{InvertMetal,InvertWater}.

Conversely, various lines of reasoning led to the introduction of
phenomenological potentials for water that are
core-softened~\cite{Others1d}. The inversion of the oxygen-oxygen
radial distribution function $g_{oo}(r)$ for water gives an effective
potential $\phi$~(Fig.~\ref{fig:InvertPotential}) that is
core-softened~\cite{InvertWater}. Thus core-softened potentials can be
considered as zeroth-order models for water.

\begin{figure}[htbp]
\begin{center}
\mbox{\psfig{figure=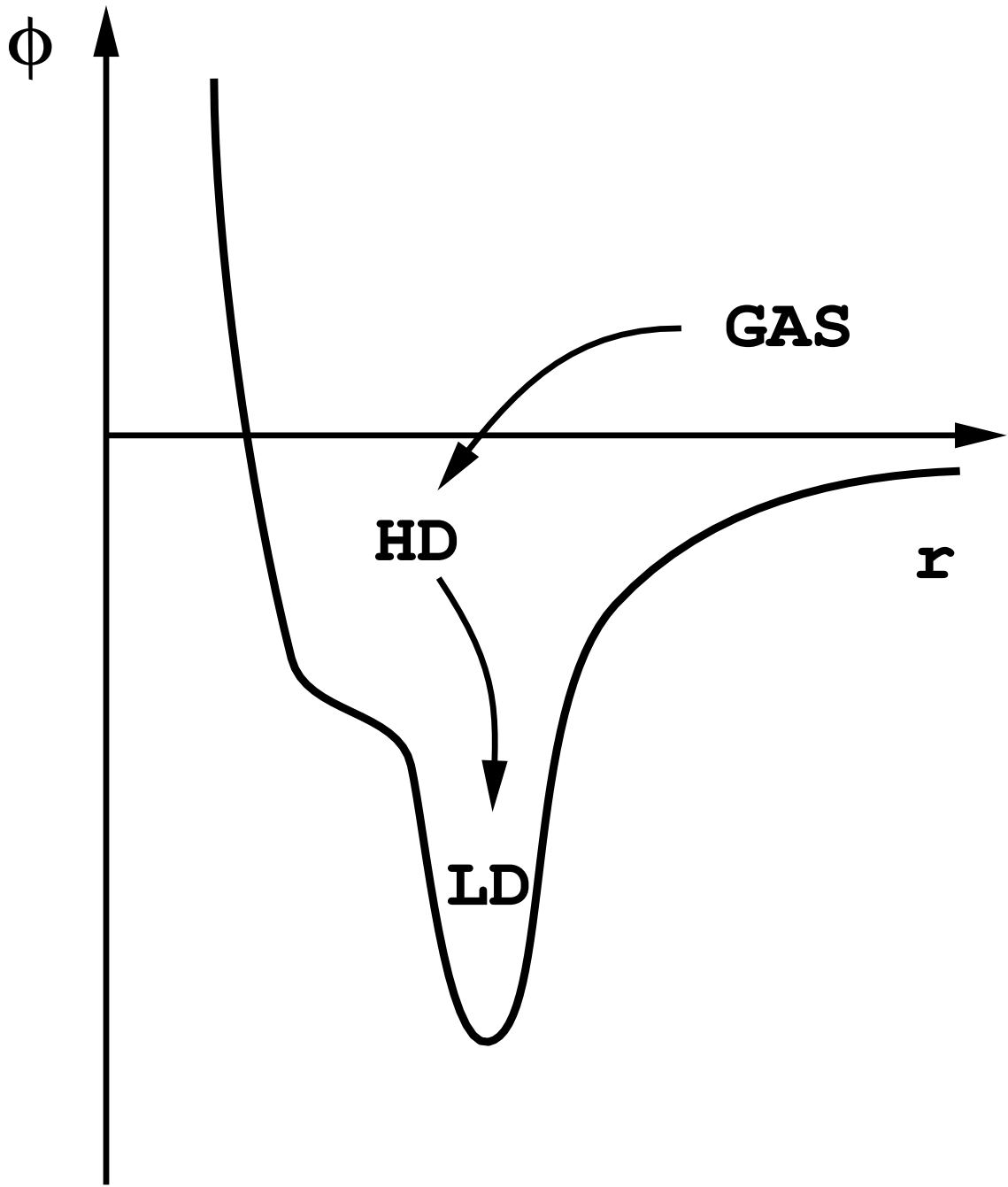,width=15cm,angle=0}}
\end{center}
\caption{On cooling from high temperatures, the system first condenses
in a liquid where the typical inter-particle distance is distributed
inside the attractive part of the potential. Further cooling causes
the distribution to be more peaked around the minimum of the
potential, making the liquid expand (HD = high density, LD = low
density).}
\label{fig:ThermalDensityAnomaly}
\end{figure}

It is natural to expect that a core-softened potential can induce a
density anomaly. In a liquid, the typical inter-particle distance is
distributed inside the attractive part of the potential. As temperature
decreases, the distribution peaks around the minimum of the potential,
causing the system to expand~(Fig.~\ref{fig:ThermalDensityAnomaly}).

In addition to the density anomaly, further studies~\cite{Sadr} of
core-softened one-dimensional fluids have discovered anomalies in the
isothermal compressibility and specific heat response functions in
water.  These anomalies are related to the existence of a critical
point at $T=0$ and high density, similar to what is conjectured to
occur in real water~\cite{Poole}.

The region where the liquid has a density anomaly must have an upper
boundary in the P-T plane; this boundary defines the line of density maxima
($T_{\mbox{\scriptsize Md}}$). The occurrence of this boundary can be
understood by first recalling that
\begin{equation}
\label{crossalphanew}
\alpha_P \propto
\langle\delta V\delta S\rangle \propto  
\left(P\langle\delta V^2\rangle 
+ \langle\delta V\delta E\rangle\right).
\end{equation}
When $\alpha_P$ is negative, the term $\langle\delta V\delta E\rangle$
must also be negative; the fluctuations that give a negative sign to
$\langle\delta V\delta E\rangle$ correspond to the regions of the
fluid where particles penetrate the softened part of the
core. However, the $\alpha_P$ anomaly must vanish at high enough
pressures where the positive $P\langle\delta V^2\rangle$ term
dominates.

Core-softened potentials with no attractive part and no liquid-gas
transition were studied in 2 and 3 dimensions by Young and Alder,
giving a P-T phase diagram with a solid-fluid coexistence line similar
to Ce or Cs~\cite{StepPotential}. The fluid-solid coexistence line has
a negatively sloped region, as in water. The phase diagram for a
core-softened potential in 2d with no attractive part has been studied
by Jagla~\cite{Jagla}, who explicitly finds a density anomaly in the
fluid region.  These results indicate that the softened part of the
core can be solely responsible for the density anomaly and suggest
that it can be related to the existence of negatively-sloped melting
lines.

\begin{figure}[htbp]
\begin{center}
\mbox{
\psfig{figure=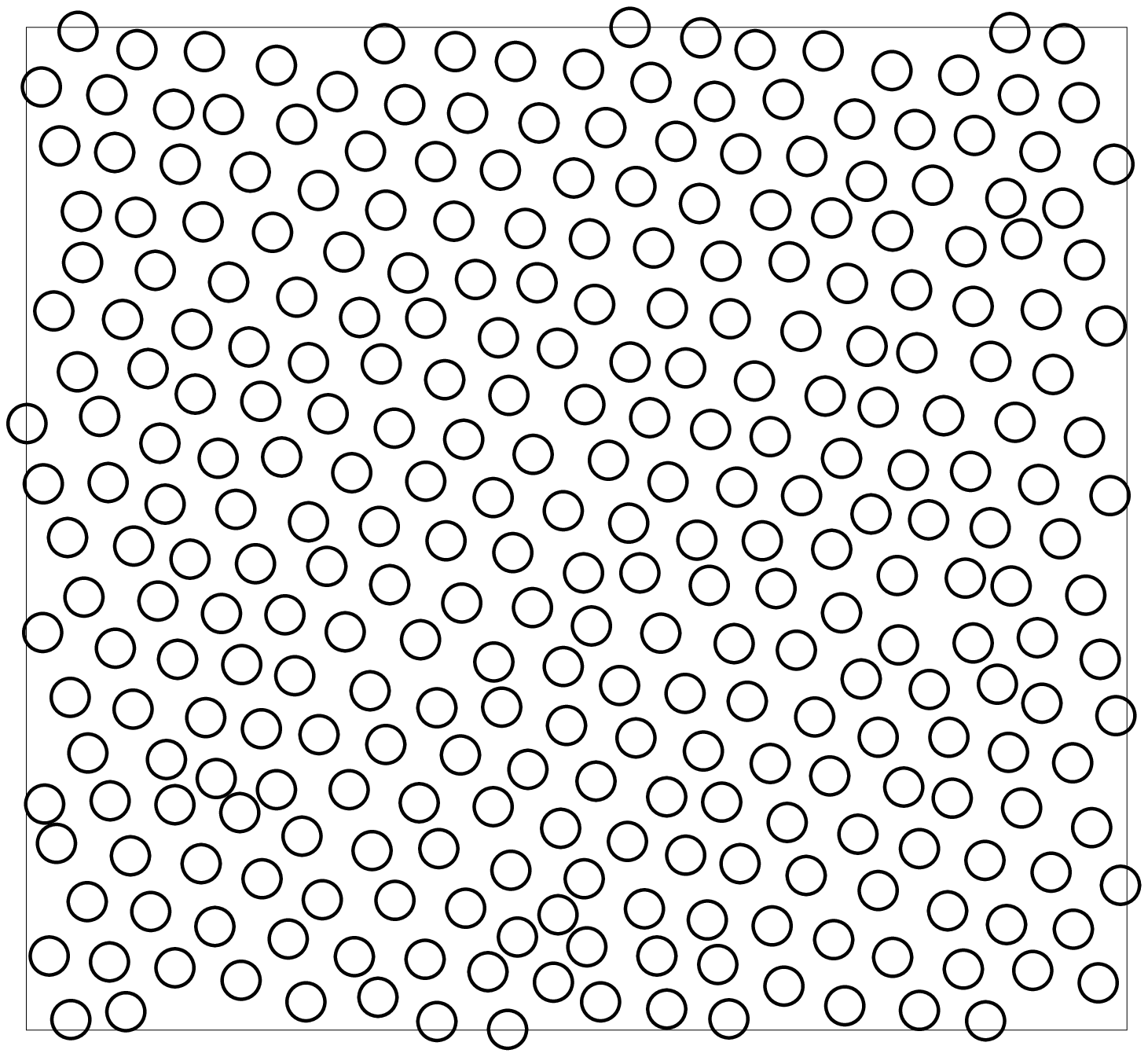,width=7cm,angle=0}
\hspace*{0.1cm}
\psfig{figure=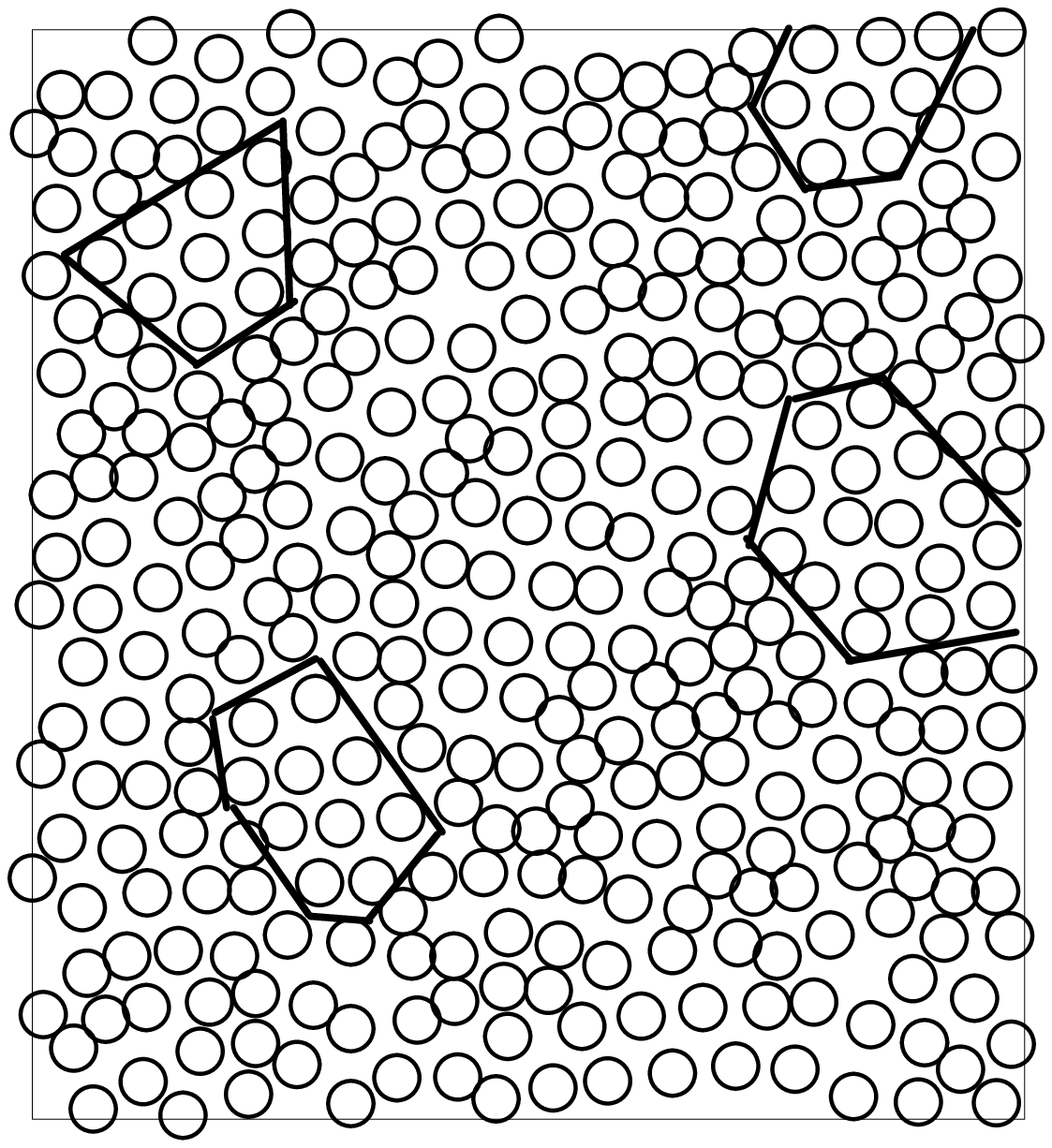,width=7cm,angle=0}
}
\end{center}
\caption{Snapshot of a low density crystal (left panel) and of a
liquid (right panel), both near the freezing line at low pressures,
for the two-dimensional core-softened model
of~\protect\cite{Sadr}. Corresponding to the fact that the freezing
line is negatively sloped, we note the appearance in the liquid of
local arrangements of the particles (black ``fences'') resembling the
low density crystal.}
\label{fig:LiquidSolid}
\end{figure}

The slope of a melting line is related through the Clausius-Clapeyron
equation $dP / dT = \Delta S / \Delta V$ to the difference $\Delta S$
in the entropies and $\Delta V$ in the volumes between the fluid and
the solid. On the other hand, the sign of the coefficient of thermal
expansion $\alpha_P \propto \langle \delta S \delta V \rangle$ depends
on the cross-correlation between entropy and volume fluctuations.
Near a melting line, one expects the relevant fluctuations in a liquid
to be ``solid-like'' as they trigger the nucleation process leading to
the first order liquid-solid transition~(Fig.~\ref{fig:LiquidSolid}).
This means that the sign of $\alpha_P\propto\langle\delta V \delta
S\rangle$ will be likely the same as $dP / dT = \Delta S/\Delta V$.

\begin{figure}[htbp]
\begin{center}
\mbox{\psfig{figure=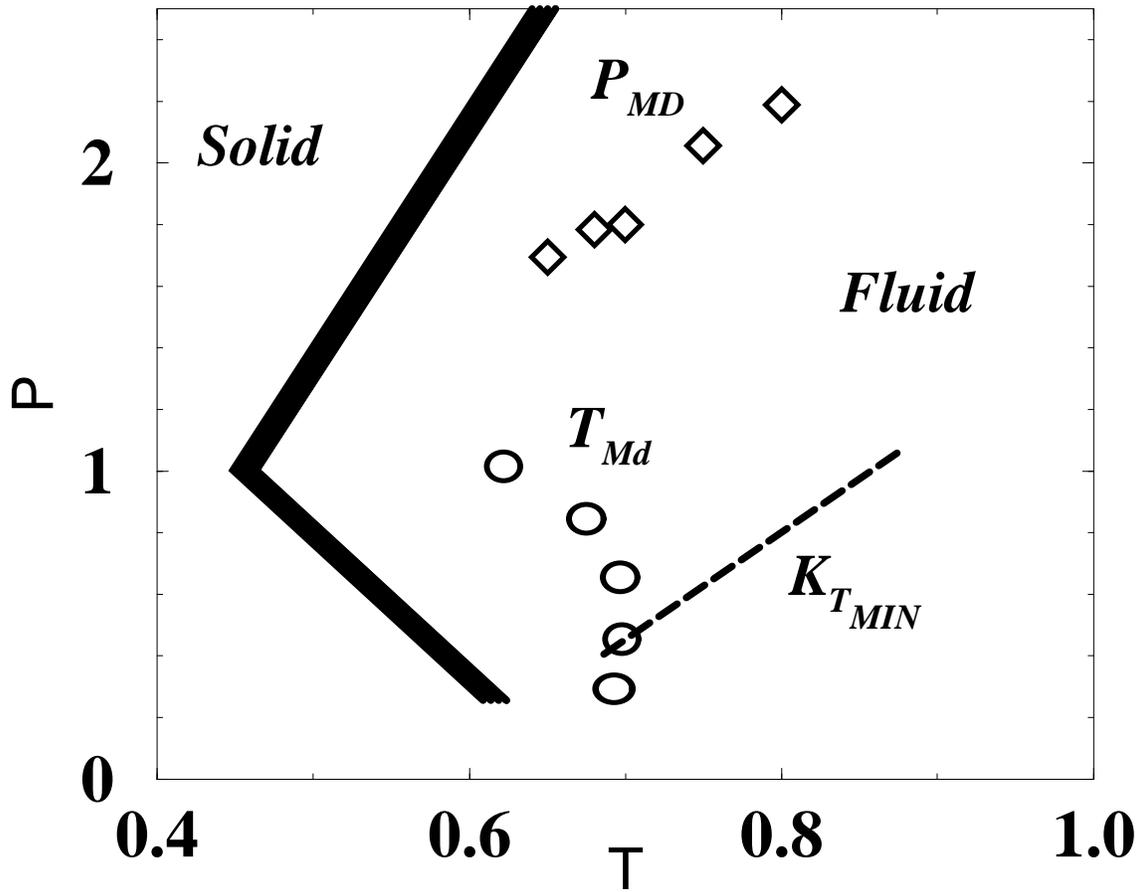,width=15cm,angle=0}}
\end{center}
\caption{Sketch of the phase diagram as found
in~\protect\cite{Sadr}. On the left of the $K_{Tmin}$ line, the
isothermal compressibility increases upon cooling. On the left of the
$T_{\mbox{\scriptsize Md}}$ line density decreases upon cooling. Below
the $P_{\mbox{\scriptsize MD}}$ line, diffusivity increases upon
pressure. Note that the density anomaly is in the region of the liquid
where the melting line is negatively sloped.}
\label{fig:PhaseDiagram}
\end{figure}

\begin{figure}[htbp]
\begin{center}
\mbox{\psfig{figure=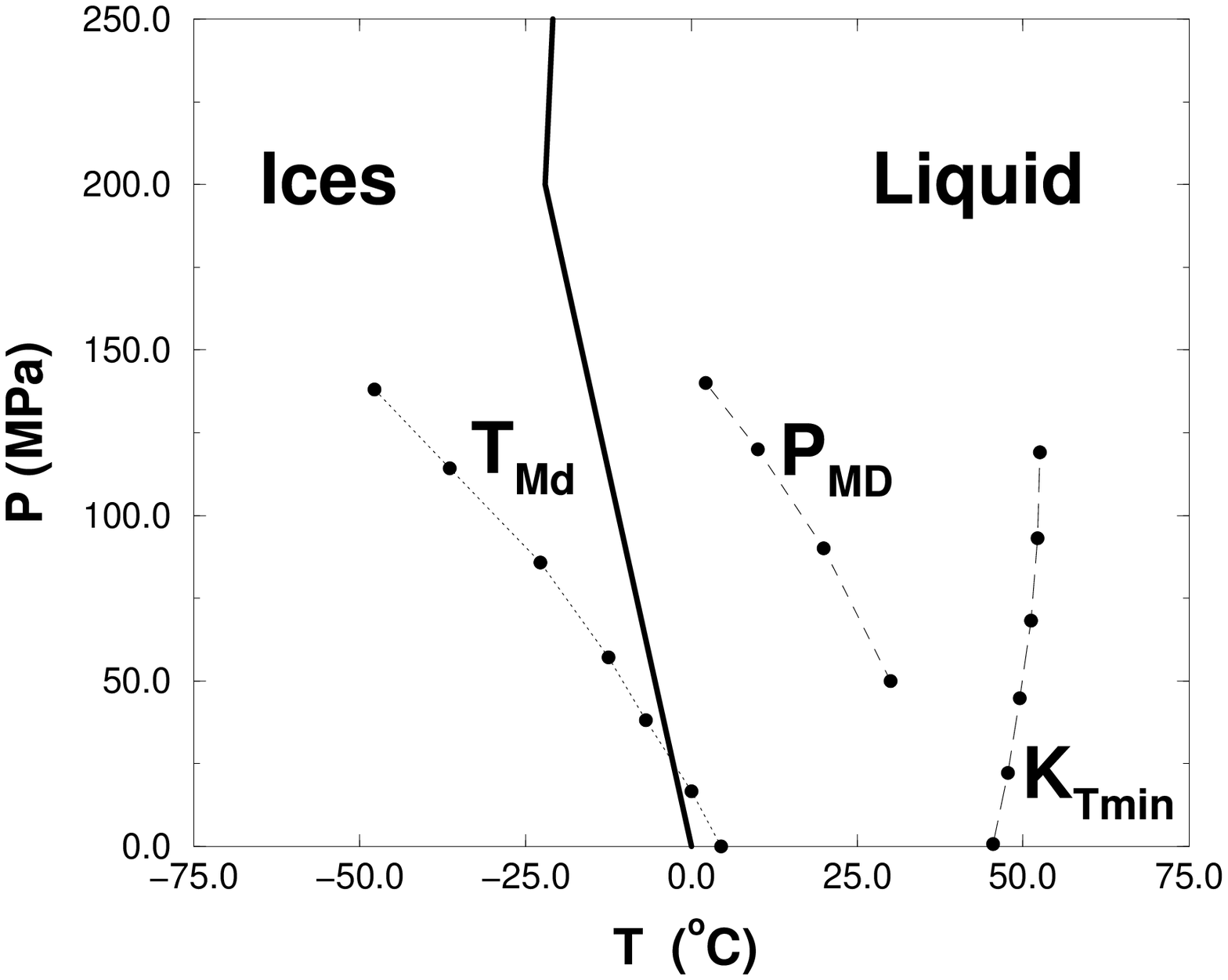,width=15cm,angle=0}}
\end{center}
\caption{Sketch of the phase diagram of water. The portions of the
$K_{Tmin}$,$T_{\mbox{\scriptsize Md}}$ and $P_{\mbox{\scriptsize MD}}$
that are beyond the melting line corresponds to experiments in the
supercooled region of water. Notice that the presence of a density
anomaly in the region of the negatively sloped melting line can occur
in the metastable phase of the liquid. Data are obtained from
refs~\protect\cite{WaterData}}.
\label{fig:PhaseWater}
\end{figure}

Extensive studies of core-softened potentials in two dimensions via
molecular dynamics simulations~\cite{Sadr} reveal a phase diagram
(Fig.~\ref{fig:PhaseDiagram}) similar to that of water
(Fig.~\ref{fig:PhaseWater})~\cite{WaterData}. Near the
negatively-sloped part of the liquid-solid freezing line ($\Delta S /
\Delta V <0$), the liquid exhibits a density anomaly ($\alpha_P
\propto \langle\delta S \delta V\rangle < 0$). In agreement with the
thermodynamic considerations of~\cite{Sastry}, the model also
reproduces the existence of a region where the isothermal
compressibility grows anomalously upon cooling in the same way as in
water. Moreover, the model succeeds in reproducing anomalies not only
in the statics, but also in the dynamics: there is a region in which
the diffusion constant anomalously increases with pressure~\cite{Sadr}
as in water. The locus of points where the diffusivity has a maximum
upon varying pressure defines the pressure of maximum diffusivity line
($P_{\mbox{\scriptsize MD}}$), which has been observed in water, in
core-softened models, and in SPC-E simulated
water~\cite{Sadr,WaterData,OurCondMatt,SPC-Efstarr}; in all these
cases the $P_{\mbox{\scriptsize MD}}$ line occurs at higher pressures
than the $T_{\mbox{\scriptsize Md}}$ line. The joint occurrence of
density and diffusivity anomalies is also observed in two-dimensional
simulations of the Gaussian-core model~\cite{Stillinger} that,
although not possessing a hard core, has a concave region in the
repulsive part of the potential.

Theories relating diffusivity to entropic
contributions~\cite{AdamGibbsDiMArzio} would predict the occurrence of
an anomaly in the diffusion $(\partial D / \partial P)_T>0$ to be
related to an anomaly in the entropy~$(\partial P / \partial
P)_T>0$. On the other hand, whenever there is a density anomaly, an
entropy anomaly occurs, as the entropy reaches a maximum along
isotherms on the $T_{\mbox{\scriptsize Md}}$ line. This is a
consequence of the Maxwell relation $(\partial S/\partial
P)_T=-(\partial V/\partial T)_P$.

In conclusion, core-softened potentials are simple realistic
potentials that can model complex fluid behavior. In addition to the
well-known liquid-gas transition, an analysis of the shape of free
energies at low temperatures reveals the presence of a second
transition, which can either be interpreted as a solid-solid or a
liquid-liquid transition. Simulations indicate that if a liquid-liquid
critical point exists it is likely to be in the region of the phase
diagram where the liquid is metastable, at least for core-softened
potentials in two dimensions~\cite{Sadr}. The presence of a
liquid-liquid critical point provides one explanation of how anomalies
in the response functions for core-softened potentials could
occur---although other scenarios are also
possible~\cite{Scenarios}. Although core-softened potentials are
capable of exhibiting most of the anomalies present in liquid water,
most materials that are known to have an effective core-softened
potential have not been studied as extensively as water, and thus the
presence of anomalies in them is still an open question.  Moreover,
the relationship between anomalies of static (e.g., entropy) and
dynamic (e.g., diffusivity) quantities is still an open issue that can
be explored using core-softened potentials, possibly within the
framework of existing theories~\cite{AdamGibbsDiMArzio}.

We thank M.~Canpolat, E.~La~Nave, M.~Meyer, S.~Sastry, F.~Sciortino,
A.~Skibinsky R.~J. Speedy, F.~W.~Starr, G.~S. Stell and D.~Wolf, for
enlighting discussions, and NSF for support.

\end{multicols}{2}

\newpage

\newpage

\newpage

\newpage

\newpage

\newpage

\newpage

\begin{thebibliography}{99}

\bibitem{Stell1} 
P. C. Hemmer and G. Stell, {\it Phys. Rev. Lett.} {\bf 24}, 1284
(1970).

\bibitem{Rowlinson} J. Rowlinson and B. Widom, {\it J. Chem. Phys.}
{\bf 52}, 1670 (1970)

\bibitem{Kac} M. Kac, G. E. Uhlenbeck, P. C. Hemmer, {\it
Journ. Math. Phys.}, {\bf 4}, 216 (1963)

\bibitem{Stell3} 
J. S. H\o ye and P. C. Hemmer, {\it Physica Norvegica\/} {\bf 7}, 1
(1973).

\bibitem{Jayaramen} A. Jayaramen, {\it Phys. Rev. A\/} {\bf 137}, 179
(1965)

\bibitem{Yoshimura} 
Y. Yoshimura, {\it Ber. Bunsenges. Phys. Chem.} {\bf 95}, 135 (1991).

\bibitem{Sadr} M.~R.~ Sadr-Lahijany, A.~Scala, S.~V.~Buldyrev and
H.~E.~Stanley, Phys.~Rev.~Lett. {\bf 81}, 4895 (1998); M. R.
Sadr-Lahijany, A. Scala, S. V. Buldyrev, and H. E.Stanley, {\it Phys.
Rev. E\/} {\bf 60}, xxx (Dec 1999); A. Scala, M. R.  Sadr-Lahijany, N.
Giovanbattista, S. V. Buldyrev, and H. E. Stanley ``Waterlike Anomalies
for Core-Softened Models of Fluids: Two Dimensions'', in preparation.

\bibitem{Stell4} 
J. M. Kincaid, G. Stell, and C. K. Hall, {\it J. Chem. Phys.} {\bf 65},
2161 (1976); J. M. Kincaid, G. Stell, and E. Goldmark, {\it J. Chem.
Phys.} {\bf 65}, 2172 (1976); J. M. Kincaid and G. Stell, {\it J. Chem.
Phys.} {\bf 67}, 420 (1977); C. K. Hall and G. Stell, Phys Rev. A {\bf
7}, 1679 (1973).

\bibitem{ExperIsostrCritPoint} 
R. I. Beecroft and C. A. Swenson, {\it J. Phys. Chem. Sol.} {\bf 15},
234 (1960); B. L. Davis and L. H. Adams, {\it J. Phys. Chem. Sol.}  {\bf
25}, 379 (1964); A. Jayaraman, {\it Phys. Rev. Sec. A\/} {\bf 137}, 179
(1965); J. M. Lawrence, M. C. Croft, and R. D. Parks, {\it
Phys. Rev. Lett.}  {\bf 35}, 289 (1975).

\bibitem{ExperCoreSoft} 
R. Sternheimer, {\it Phys. Rev.} {\bf 78}, 235 (1950); T. H. Hall,
L. Merril, and J. D. Barnett, {\it Science\/} {\bf 146}, 1297 (1964);
A. Jayaraman, {\it Phys. Rev.} {\bf 159}, 527 (1967); A. Jayaraman, {\it
Ann. Rev. Mat. Sci.} {\bf 2}, 121 (1972); M. B. Maple, and D. Wohlleben,
{\it AIP Conf. Proc.} {\it 18}, 447 (1974); P. W. Anderson and
S. T. Chui, {\it Phys. Rev. B\/} {\bf 9}, 3229 (1975); A. Jayaraman,
P. Dernier, and L. D. Longinotti, {\it Phys. Rev. B\/} {\bf 11}, 2783
(1975); A. W. Lawson and Ting-Yuan Tang, {\it Phys. Rev.} {\bf 76}, 301
(1949); A. F. Schuch and J. H. Sturdivant, {\it J. Chem. Phys.} {\bf
18}, 145 (1950); C. J. McHargue and H. Y. Yakel Jr., {\it Acta Metall.}
{\bf 8}, 637 (1960); M. Wilkinson, H. Child, C. McHargue, W. Koehler,
and F. Wollan,{\it Phys. Rev.} {\bf 122}, 1409 (1961); R. Ramirez and
L. M. Falivov, {\it Phys. Rev. B\/} {\bf 3}, 2425 (1975); L. F. Bates
and M. M. Newmann, {\it Proc. Phys. Soc. London\/} {\bf 72}, 345 (1958).

\bibitem{CalculateMetal} 
K. K. Mon, N. W. Ashcroft, and G. V. Chester, {\it Phys. Rev. B\/} {\bf
19}, 5103 (1979).

\bibitem{InvertMetal} 
I. Yokoyama and S. Ono, {\it J. Phys. F: Met. Phys.} {\bf 15}, 1215
(1985); K. Hoshino, C. H. Leung, I. L. McLaughlin, S. M. M. Rahman, and
W. H. Young, {\it J. Phys. F: Met. Phys.} {\bf 17}, 787 (1987).

\bibitem{Stell2} 
G. Stell and P. C. Hemmer, {\it J. Chem. Phys.} {\bf 56}, 4274 (1972).

\bibitem{Debenedetti}
P. G. Debenedetti, { \it Metastable Liquids\/} (Princeton University
Press, Princeton, 1996); P.~G.~Debenedetti, V. S. Raghavan and S. S.
Borick, J.~Phys.~Chem. {\bf 95}, 4540 (1991); P. G. Debenedetti and M.
C. Dantonio, AICHE J. {\bf 34}, 447 (1988).

\bibitem{Takahashi} H.~Takahashi, {\it Proc. Phys. Math. Soc. Japan},
{\bf 24}, 60 (1942); {\it Mathematical Physics in One Dimension},
edited by E.~H.~Lieb and D.~C.~Mattis (Academic, New York, 1966)
pp. 25-34.


\bibitem{InvertWater} 
T. Head-Gordon and F. H. Stillinger, {\it J. Chem. Phys.} {\bf 98}, 3313
(1993).

\bibitem{Others1d} 
A. Ben-Naim, {\it Statistical Thermodynamics for Chemists and
Biochemists\/} (Plenum Press, New York, 1992), pp.~233--238; C.~H.~Cho
et al., Phys. Rev. Lett.  {\bf 76}, 1651 (1996); M.  Canpolat, F. W.
Starr, A. Scala, M. R.  Sadr-Lahijany, O.  Mishima, S.  Havlin and H. E.
Stanley, Chem. Phys. Lett. {\bf 294}, 9 (1998).

\bibitem{Poole} 
P. H. Poole, F. Sciortino, U. Essman, and H. E. Stanley, {\it Nature \/}
{\bf 360}, 324 (1992); Phys. Rev. E {\bf 48}, 4605 (1993); O. Mishima and
H. E. Stanley, Nature {\bf 392}, 164 (1998).

\bibitem{StepPotential} 
D. A. Young and B. J. Alder, {\it Phys. Rev. Lett.}  {\bf 38}, 1213
(1977); D. A. Young and B. J. Alder, {\it J. Chem. Phys.} {\bf 70}, 473
(1979).

\bibitem{Jagla} 
E. A. Jagla, {\it Phys Rev E\/} {\bf 58}, 1478 (1998).

\bibitem{WaterData} F.X. Prielmeier, E.W. Lang, R.J. Speedy, and
H.-D. L\"udemann, Phys. Rev. Lett. {\bf 59}, 1128 (1987);
Ber. Bunsenges. Phys. Chem. {\bf 92}, 1111 (1988); L. Haar,
J.S.Gallagher, G.S. Kell, ``NBS/NRC Steam Tables. Thermodynamic and
Transport Properties and Computer Programs for Vapor and Liquid States
of Water in SI Units'', Hemisphere Publishing Co., Washington,
D.C. 271-276 (1984).

\bibitem{Sastry} S.~Sastry et al., Phys.~Rev.~E {\bf 53}, 6144 (1996).

\bibitem{OurCondMatt} 
A. Scala, F. W. Starr, E. La Nave, F. Sciortino, and H. E. Stanley,
``Configurational Entropy and Diffusivity of Supercooled Water,''
cond-mat/9908301; E. La Nave, A. Scala, F. W. Starr, F. Sciortino, and
H. E. Stanley, ``Instantaneous Normal Mode Analysis of Supercooled
Water,'' cond-mat/9908412.

\bibitem{SPC-Efstarr} F.W.~Starr, S.Harrington F~.Sciortino and
 H.E.~Stanley Phys.~Rev.~Lett. {\bf 82}, 3629 (1999); F.W.~Starr,
 F~.Sciortino and H.E.~Stanley Phys.~Rev.~E {\bf 60}, XXX (1999)

\bibitem{Stillinger}
F.~H.~Stillinger and D.~K.~Stillinger, Physica~A {\bf 244}, 358 (1997);
F.~H.~Stillinger and T. A. Weber, J.~Chem.~Phys.  {\bf 68}, 3837 (1978);
{\it Ibid.} {\bf 74}, 4015 (1981).

\bibitem{AdamGibbsDiMArzio}
G. Adam and J. H. Gibbs, J. Chem. Phys. {\bf 43}, 139 (1965).

\bibitem{Scenarios} 
O. Mishima and H. E. Stanley, {\it Nature\/} {\bf 396}, 329 (1998).

\end{thebibliography}
\end{document}